# Recent Results on the Nucleon Structure Functions from Lattice QCD[a]

M. Göckeler[1,2], R. Horsley[3], E.-M. Ilgenfritz[3], H. Oelrich[1], H. Perlt[4], P. Rakow[1],
G. Schierholz[5,1] and A. Schiller[4]

[1] *Gruppe Theorie der Elementarteilchen, Höchstleistungsrechenzentrum HLRZ,
c/o Forschungszentrum Jülich, D-52425 Jülich, Germany*

[2] *Institut für Theoretische Physik, RWTH Aachen, D-52056 Aachen, Germany*

[3] *Institut für Physik, Humboldt-Universität, D-10115 Berlin, Germany*

[4] *Institut für Theoretische Physik, Universität Leipzig, D-04109 Leipzig, Germany*

[5] *Deutsches Elektronen-Synchrotron DESY, D-22603 Hamburg, Germany*

We report on recent results of a high statistics lattice calculation of the unpolarized and polarized structure functions of the nucleon.

## 1 Introduction

Our knowledge of the deep-inelastic structure functions of the nucleon, both for unpolarized and polarized beams and targets, has improved a lot in the last couple of years [1]. But all efforts of our experimental colleagues are in vain if we do not succeed in computing the distribution functions from first principles.

This is basically a nonperturbative problem. The tool to solve it is lattice gauge theory. We have seen [2,3] that this is possible with the help of dedicated computers, and in the next couple of years we hope to make real progress in this endeavor.

The theoretical basis of the calculation is the operator product expansion, which relates the moments of the structure functions to forward nucleon matrix elements of local operators:

$$\int_0^1 dx\, x^{n-2} F_2(x, Q^2)$$
$$= \sum_{f=u,d,\cdots} c_{2,n}^{(f)}(\mu^2/Q^2, g(\mu)) v_n^{(f)}(\mu),$$
$$2 \int_0^1 dx\, x^n g_1(x, Q^2)$$
$$= \frac{1}{2} \sum_{f=u,d,\cdots} e_{1,n}^{(f)}(\mu^2/Q^2, g(\mu)) a_n^{(f)}(\mu), \quad (1)$$
$$2 \int_0^1 dx\, x^n g_2(x, Q^2)$$
$$= \frac{1}{2} \frac{n}{n+1} \sum_{f=u,d,\cdots} [e_{2,n}^{(f)}(\mu^2/Q^2, g(\mu)) d_n^{(f)}(\mu)$$
$$- e_{1,n}^{(f)}(\mu^2/Q^2, g(\mu)) a_n^{(f)}(\mu)].$$

The Wilson coefficients are computed perturbatively, and the reduced matrix elements $v_n$, $a_n$ and $d_n$, which are defined by

$$\frac{1}{2} \sum_{\vec{s}} \langle \vec{p}, \vec{s} | \mathcal{O}_{\{\mu_1 \cdots \mu_n\}}^{(f)}(\mu) | \vec{p}, \vec{s} \rangle$$
$$= 2 v_n^{(f)}(\mu) [p_{\mu_1} \cdots p_{\mu_n} - \text{traces}],$$
$$\langle \vec{p}, \vec{s} | \mathcal{O}_{\{\sigma \mu_1 \cdots \mu_n\}}^{5(f)}(\mu) | \vec{p}, \vec{s} \rangle$$
$$= \frac{1}{n+1} a_n^{(f)}(\mu) [s_\sigma p_{\mu_1} \cdots p_{\mu_n} + \cdots], \quad (2)$$
$$\langle \vec{p}, \vec{s} | \mathcal{O}_{[\sigma \{\mu_1] \cdots \mu_n\}}^{5(f)}(\mu) | \vec{p}, \vec{s} \rangle$$
$$= \frac{1}{n+1} d_n^{(f)}(\mu) [(s_\sigma p_{\mu_1} - s_{\mu_1} p_\sigma) p_{\mu_2} \cdots p_{\mu_n} + \cdots]$$

($\{\cdots\}$ and $[\cdots]$ denote symmetrization and antisymmetrization, respectively) with

$$\mathcal{O}_{\mu_1 \cdots \mu_n}^{(f)} = \left(\frac{i}{2}\right)^{n-1} \bar\psi^{(f)} \gamma_{\mu_1} \overleftrightarrow{D}_{\mu_2} \cdots \overleftrightarrow{D}_{\mu_n} \psi^{(f)} - \text{traces},$$
$$\mathcal{O}_{\sigma \mu_1 \cdots \mu_n}^{5(f)} = \left(\frac{i}{2}\right)^n \bar\psi^{(f)} \gamma_\sigma \gamma_5 \overleftrightarrow{D}_{\mu_1} \cdots \overleftrightarrow{D}_{\mu_n} \psi^{(f)} - \text{traces}, \quad (3)$$

are computed on the lattice. Here we have restricted ourselves to the leading quark operators. The structure functions are independent of the subtraction point $\mu$. The moments have the parton model interpretation:

$$v_n^{(f)} = \langle x^{n-1} \rangle^{(f)}, \ a_0^{(u)} = 2\Delta u, a_0^{(d)} = 2\Delta d, \quad (4)$$

---

[a]Talk given by G. Schierholz at *International Europhysics Conference on High Energy Physics*, Brussels, July 27 - August 2, 1995

where $x$ is the fraction of the nucleon momentum carried by the quarks and $\Delta u, \Delta d$ denote the quark spin contribution to the spin of the nucleon. The structure function $g_2$ consists of two contributions: $a_n^{(f)}$ has twist two, whereas $d_n^{(f)}$ has twist three. The latter contribution has no parton model interpretation.

## 2 Highlights of the Calculation

The (euclidean) lattice operators are obtained from (3) by the substitution

$$\frac{i}{2}\overleftrightarrow{D}_\mu \rightarrow D_\mu, \quad (5)$$

where $D_\mu$ is the lattice covariant derivative. Choosing suitable interpolating fields $B, \bar{B}$ for the nucleon, the desired nucleon matrix elements can be derived from ratios of three-point to two-point functions,

$$\frac{\langle B(t)\mathcal{O}(\tau)\bar{B}(0)\rangle}{\langle B(t)\bar{B}(0)\rangle} = \langle N|\mathcal{O}|N\rangle + \cdots \quad (6)$$

for $0 \ll \tau \ll t$.

The calculations in this talk are done on $16^3 \times 32$ lattices at $\beta = 6.0$. We use Wilson fermions and work in the quenched approximation, where the effect of dynamical quark loops has been neglected. To be able to extrapolate our results to the chiral limit, and to make contact to the predictions of the quark model, we do the calculations at three different hopping parameters, $\kappa = 0.155, 0.153$ and $0.1515$, which correspond to physical quark masses of roughly $m_q = 70, 130$ and $190$ MeV, respectively. So far we have analyzed of the order of 1000, 800 and 400 independent gauge field configurations, respectively, at the three values of $\kappa$.

The chiral limit is obtained by extrapolating $m_\pi^2$ linearly to zero in $1/\kappa$. For the critical hopping parameter we find $\kappa_c = 0.15693(4)$. At our smallest quark mass, taking the nucleon mass as the scale, the inverse lattice spacing is $a^{-1} \approx 1.4$ GeV, which corresponds to a distance of 0.14 fm and a spatial box size of 2.2 fm. In the chiral limit the inverse lattice spacing varies between 1.8 and 2.2 GeV, depending on how we extrapolate the nucleon mass to the chiral limit.

The operators must be chosen such that they belong to an irreducible representation of the hypercubic group $H(4)$. We consider the operators

| Operators | $\langle\mathcal{O}\rangle$ |
|---|---|
| $\mathcal{O}_{\{44\}} - \frac{1}{3}(\mathcal{O}_{\{11\}} + \mathcal{O}_{\{22\}} + \mathcal{O}_{\{33\}})$ | $v_2$ |
| $\mathcal{O}_{\{114\}} - \frac{1}{2}(\mathcal{O}_{\{224\}} + \mathcal{O}_{\{334\}})$ | $v_3$ |
| $\mathcal{O}_{\{1144\}} + \mathcal{O}_{\{2233\}} - \mathcal{O}_{\{1133\}} - \mathcal{O}_{\{2244\}}$ | $v_4$ |
| $\mathcal{O}_2^5$ | $a_0$ |
| $\mathcal{O}_{\{214\}}^5$ | $a_2$ |
| $\mathcal{O}_{[2\{1]4\}}^5$ | $d_2$ |

They require at most one spatial component of the nucleon's momentum to be non-zero [2].

The lattice operators must be brought into a continuum renormalization scheme, e.g. the momentum subtraction scheme or the $\overline{MS}$ scheme. Schematically this is written

$$\mathcal{O}(\mu) = Z_\mathcal{O}(a\mu)\mathcal{O}(a). \quad (7)$$

We have computed the renormalization constants $Z_\mathcal{O}$ perturbatively to one loop order [2] and non-perturbatively [4] following a recent suggestion in the literature [5]. The non-perturbative calculation uses $O(100)$ gauge-fixed lattice gauge field configurations. In Fig. 1 we compare both results for some of our operators. We find that for larger values of $\mu^2$ perturbative and non-perturbative renormalization constants agree within 20% and better. At smaller values of $\mu^2$ we cannot really expect to find agreement with perturbation theory yet at our value of $\beta$ because here the internal momenta are no longer large compared to any hadronic scale. In the following we have taken the perturbative values for the renormalization constants. The numbers we will quote are for $\mu = a^{-1}$.

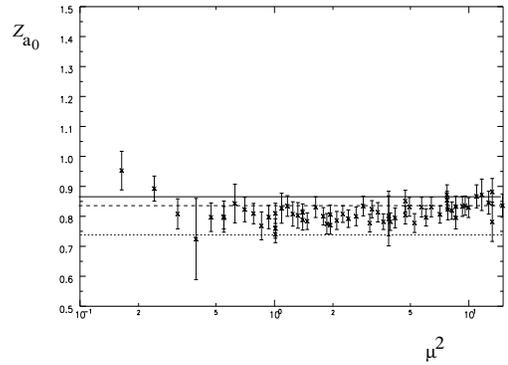

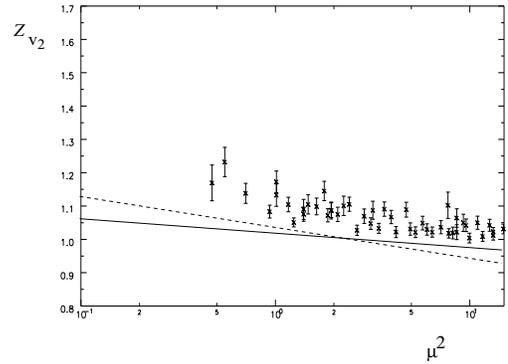

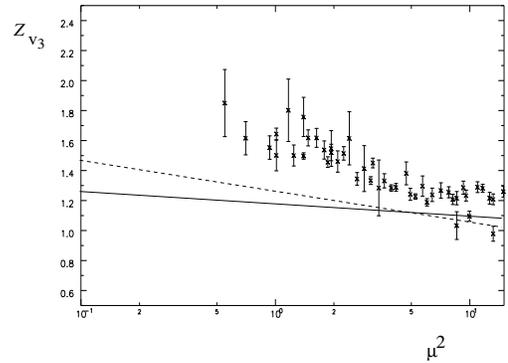

Figure 1: The renormalization constants $Z_{a_0}, Z_{v_2}$ and $Z_{v_3}$ as a function of $\mu^2$ (in lattice units) for $\kappa = 0.1530$. The solid (dotted and dashed) lines are the results of standard (tadpole improved) perturbation theory as discussed in Ref. 3.

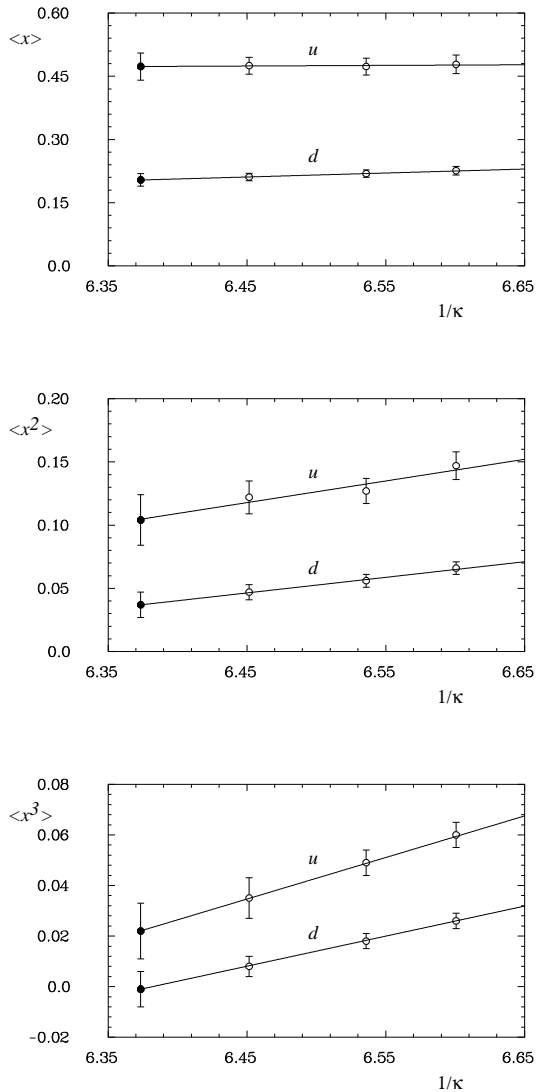

Figure 2: The moments $\langle x \rangle$, $\langle x^2 \rangle$ and $\langle x^3 \rangle$ as a function of $1/\kappa$.

## 3  Unpolarized Structure Functions

Let us first discuss our results for the unpolarized structure functions. The moments $\langle x^{n-1} \rangle$ are shown in Fig. 2 for our three quark masses. The lines are linear fits to the data, and the solid circles are the results of the extrapolation to the chiral limit. The lowest moment is found to be practically independent of the quark mass, while with increasing $n$ the moments show a growing increase with the quark mass.

One is tempted to compare the quenched quark distribution functions with the phenomenological valence quark distribution functions. But this is not quite correct because we are dealing with probabilities. Consider $\langle x \rangle$, and let the subscript $V$, $S$ and $G$ denote the valence quark, sea quark and gluon contribution. Momentum conservation then demands that $\langle x \rangle_V^{(u)} + \langle x \rangle_V^{(d)} + \langle x \rangle_S + \langle x \rangle_G = 1$. In the quenched approximation $\langle x \rangle_S = 0$, so that valence quark and gluon contributions should sum up to one. If, for example, the gluon contribution is approximately equal in both cases, we would thus expect to find that the quenched quark result is noticeably larger than the valence quark contribution. A better quantity to compare is the difference of $u$- and $d$-quark distribution functions. In the chiral limit we obtain $\langle x \rangle^{(u)} - \langle x \rangle^{(d)} = 0.26(4)$. Martin, Roberts and Stirling [6] (fit $D_-$) find 0.18, while the CTEQ collaboration [7] (fit CTEQ3M) obtains 0.22. Our result is one standard deviation away from the CTEQ result. For the higher moments we find agreement with the phenomenological numbers within the error bars.

## 4  Polarized Structure Functions

We shall now turn to the disscussion of the polarized structure functions. Let us first concentrate on $\Delta u$ and $\Delta d$. In Fig. 3 we show our results together with a linear fit to the data. The solid circles indicate the extrapolation to the chiral limit. For heavy quark masses we find $\Delta u \approx 1$ and $\Delta d \approx -1/4$, in good agreement with the quark model [8]. In the chiral limit we obtain

$$\Delta u - \Delta d \equiv g_A = 1.07(9). \qquad (8)$$

This difference may be expected to be insensitive to the effects of dynamical quarks. The experimental value for the axial vector coupling constant is $g_A = 1.26$. Our result and the experimental value differ by two standard deviations.

If we combine this result with the perturbatively known Wilson coefficients we obtain for the Bjorken integral

$$\int_0^1 dx (g_1^p(x, Q^2) - g_1^n(x, Q^2)) = 0.174(25) \qquad (9)$$

at $Q^2 = a^{-2} \approx 3 - 5\,\text{GeV}^2$. Because of uncertainties in the mass extrapolations to the chiral limit as mentioned before, we cannot state the scale more precisely at the moment. Our result (9) is in agreement with the experimental value [9] of 0.163(17). For the second moment of $g_1$ we find

$$\int_0^1 dx\, x^2 (g_1^p(x,Q^2) - g_1^n(x,Q^2)) = 0.0162(54). \quad (10)$$

This number is also consistent with experiment.

Let us finally discuss the structure function $g_2$. An interesting question is how large the contribution of the twist-three operator is. In Fig. 4 we have shown $d_2$ for our three values of the quark mass together with a linear fit. In the limit of heavy quark masses $d_2$ goes to zero as one would have expected. In the chiral limit, on the other hand, $d_2$ is compatible in magnitude to $a_2$ (which we have no space to show here [4]). Altogether we obtain

$$\int_0^1 dx\, x^2 (g_2^p(x,Q^2) - g_2^n(x,Q^2)) = -0.0257(60). \quad (11)$$

Without the twist-three contribution the number would have been $-0.0109(35)$.

## 5 Conclusions and Outlook

We have seen that an accurate calculation of the structure functions of the nucleon from first principles is feasible. Our results so far are encouraging. If they do not exactly reproduce the experimental values this may have various reasons.

So far we have only considered the leading quark operators and neglected the contributions of gluonic operators. Gluonic operators are much harder to get at. A preliminary study of matrix elements involving gluonic operators shows however that it is possible to compute the gluon distribution functions accurately as well. We hope to present first results in the near future.

For a quantitative analysis at low $Q^2$ it will also be important to include the contribution of higher twist operators. As we have seen in the case of $d_2$, higher twist

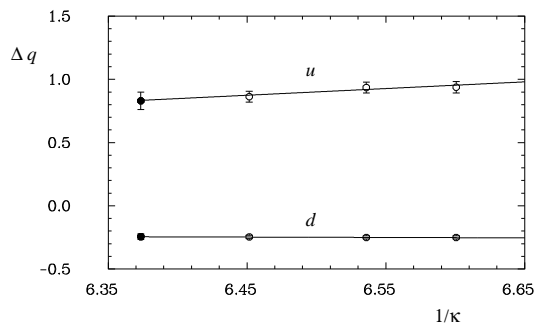

Figure 3: The quark spin contributions to the spin of the nucleon $\Delta u$ and $\Delta d$ as a function of $1/\kappa$.

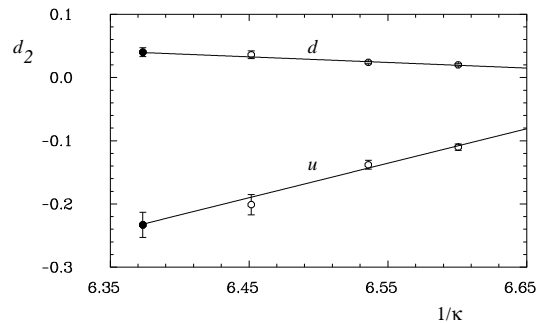

Figure 4: The twist-three matrix element $d_2$ as a function of $1/\kappa$.

contributions may be large, contrary to one's expectations.

Other sources of uncertainties are finite cut-off and volume effects as well as the lack of dynamical quark loops. Certain tests of finite cut-off effects have shown that they are presumably small [2]. We know how these uncertainties can be systematically reduced, and we will do so in the future.

## Acknowledgments

This work is supported in part by the Deutsche Forschungsgemeinschaft. The calculations are done on the Quadrics computers at DESY-Zeuthen and the University of Bielefeld. We thank both institutions for their support.